\documentclass[11pt,twocolumn,showpacs,floatfix,superscriptaddress,nofootinbib,amssymb,amsmath,aps,prc]{revtex4-2}

\usepackage{CJK} 
\usepackage[dvips]{graphicx}
\usepackage{epstopdf} 
\usepackage{latexsym,amssymb,amsmath,epsfig,bm,times,psfrag,subfig}
\usepackage{color}
\usepackage{mathptmx}                
\usepackage{dcolumn}                 
\usepackage[bookmarksnumbered,bookmarksopen,colorlinks,citecolor=blue,linkcolor=blue]{hyperref}
\renewcommand{\part}{{\rm part}}

\renewcommand{\bm}{\boldsymbol}
\newcommand{\be}{\begin{equation}}
\newcommand{\ee}{\end{equation}}
\newcommand{\bear}{\begin{eqnarray}}
\newcommand{\eear}{\end{eqnarray}}
\newcommand{\ba}{\begin{array}}
\newcommand{\ea}{\end{array}}

\begin{document}
	\begin{CJK*}{UTF8}{gbsn}

	\title{A new method to clarify contribution of chiral magnetic effect in small collision system $p^{\uparrow} + A$ involving a transversely polarized proton }

	\author{Gui-Zhen Wu (吴桂珍)}
	\thanks{These authors contribute equally to this work.}
	\affiliation{School of Physics, Huazhong University of Science and Technology, Wuhan 430074, China}
	
	\author{Zong-Wei Zhang (张宗炜)}
	\thanks{These authors contribute equally to this work.}
	\affiliation{Modern Industrial School of Health Management, Jinzhou Medical University, Jinzhou 121000, China}
	\affiliation{School of Physics, Huazhong University of Science and Technology, Wuhan 430074, China}

	\author{Chen Gao (高晨)}
	\affiliation{School of Physics, Huazhong University of Science and Technology, Wuhan 430074, China}

	\author{Yi Xu (许易)}
	\affiliation{School of Physics, Huazhong University of Science and Technology, Wuhan 430074, China}
	
	\author{Wei-Tian Deng (邓维天)}
	\thanks{The corresponding author}
	\email{dengwt@hust.edu.cn}
	\affiliation{School of Physics, Huazhong University of Science and Technology, Wuhan 430074, China}


\begin{abstract}
	In this paper, we propose a new experiment method to check contribution of chiral magnetic effect (CME). With experimental data of DIS involving transversely polarized proton, we have calculated the 3-D charge density inside the polarized proton, which is found to have a significant violation to spherical symmetry.  Then we have calculated the property of electromagnetic field (E-M field) generated by a single transversely polarized proton ($p^{\uparrow}$). Based on them, the E-M field generated in small collision system $p^{\uparrow}+A$  are studied. We find that the orientation of this  E-M field has a significant dependence on the polarization direction of  the proton, and the correlator ($\Delta\gamma$ ) has also significant dependence on the angle between reaction plane and polarization direction. As background contribution are canceled comparing two collision geometry schemes, only contribution of CME is remained. 
\end{abstract}
\maketitle
\end{CJK*} 

In recent years, it has been recognized that there may be a CP violation effect\cite{Kharzeev:1998kz,Kharzeev:2004ey,Kharzeev:2007jp,Fukushima:2008xe,Kharzeev:2015znc,Liu:2020ymh} due to the quantum anomalous fluctuation of  QCD vacuum.  This effect can be observed as chiral magnetic effect (CME) in quark-gluon plasma (QGP) with existing a strong magnetic fields\cite{Kharzeev:2007jp,Fukushima:2008xe}.  In high energy heavy ion collision experiment, QGP is produced since the extreme environment needed for phase transition is satisfied, and a strong magnetic field is also produced in non-central heavy ion collision \cite{Rafelski:1975rf,Skokov:2009qp,Bzdak:2011yy,Voronyuk:2011jd,Deng:2012pc,Deng:2014uja,Inghirami:2016iru,Yan:2021zjc}. 
So heavy ion collision is an excellent place to study CME. A charge asymmetry (charge separation) along magnetic field is induced by the CME, which provide a way to study the quantum anomaly of QCD vacuum topology.

In order to study the  charge separation, a three-point correlator $\gamma$ is introduced\cite{Voloshin:2004vk}
\begin{equation}
	\gamma_{\alpha\beta}=< \cos (\Phi_{\alpha}+\Phi_{\beta}-2\Phi_{RP})>
\end{equation}
Where $\Phi_{\alpha}$ and $\Phi_{\beta}$  are azimuthal angles of charged particle $\alpha$ and $\beta$,  $\Phi_{RP}$  is the azimuthal angle of  reaction plane respective to Lab frame. To cancel charge-independent backgrounds, the difference of correlator $\Delta \gamma = \gamma_{OS} - \gamma_{SS}$ is used\cite{Kharzeev:2015znc}, where  $\gamma_{OS}$ represents for opposite charge particle pair, $\gamma_{SS}$ represents for same charge particle pair.  The contribution of CME into correlator can be expressed as\cite{Bloczynski:2012en,Bloczynski:2013mca}
\begin{equation}
	\Delta\gamma_\mathrm{CME} \propto B^2\cdot\cos [2(\Phi_B-\Phi_{RP})],
	\label{Eq-Delta-gamma-PhiRP}
\end{equation} 
where $\Phi_B$ is the azimuthal angle of magnetic field. 
However,  a large mount of QCD background effects are also included in  $\Delta \gamma$ which is related to QGP flow $v_2$ \cite{Xu:2017zcn,Adamczyk:2013kcb,Wang:2016iov,Ajitanand:2010rc,Bzdak:2011np,Zhao:2017nfq}. 
 It's a huge challenge to clarify the contributions of CME and background correspondingly in experiment data of $\Delta \gamma_\mathrm{exp}$.
 \begin{equation}
	\Delta\gamma_\mathrm{exp}  = \Delta\gamma_\mathrm{CME}  + \Delta\gamma_\mathrm{BG}.
	\label{Eq-Delta-gamma-CME-BG}
\end{equation} 
In this paper, we propose a new method to check and clarify the contribution of CME in small collision system involving a transversely polarized proton $p^{\uparrow} + A$

In our previous work\cite{Zhang:2021jrc}, employing a physical Charge-Profile model to describe the inner charge distribution of a proton,  we calculated the property of electromagnetic field produced in small system p+A. We find that there is a significant azimuthal correlation between $\Phi_B$ and $\Phi_{RP}$.  Here, we forward our study to  transversely polarized proton.

\begin{figure}
	\includegraphics[width=0.8\linewidth]{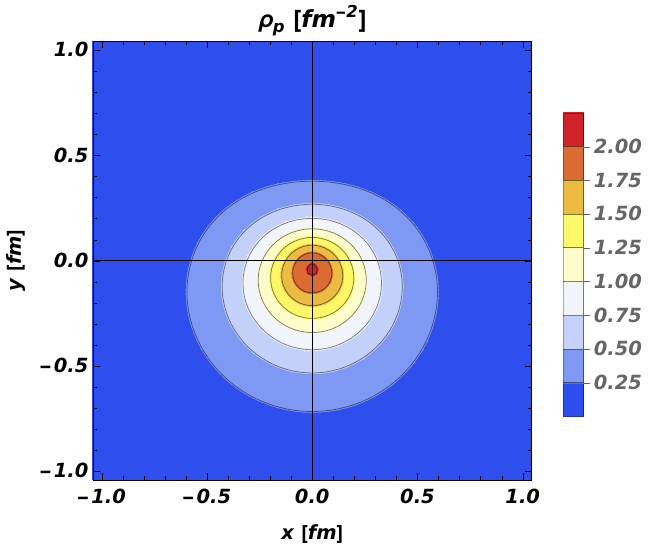}
	\caption{Plane charge density in a proton polarized along $+x$ direction.   }
		\label{fig-rho2D-polarzied-p}
\end{figure}

In DIS onto transversely polarized nucleon experiment\cite{Carlson:2007xd}, the plane charge distribution of a polarized proton is given
\begin{align}
	\rho_T^N(\bm{b})&=\rho_0^N(b)  -\sin(\phi_b-\phi_S)   \nonumber   \\
	&\int\limits_0^{\infty}\frac{\rm{d} Q}{2\pi}\frac{Q^2}{2M_N}J_1(bQ)\frac{G_M(Q^2)-G_E(Q^2)}{1+\tau}.
\end{align}
The first term in it is the  plane charge distribution of unpolarized proton, which is also given by DIS experiment \cite{Miller:2010nz}
	\begin{align}
	     \rho_0^N(\bm{b})=\int\limits_0^{\infty}\frac{\rm{d} Q}{2\pi}Q J_0(bQ)  \frac{G_E(Q^2)+\tau G_M(Q^2)}{1+\tau}.
    \end{align}
where $\tau=\frac{Q^2}{4M^2}$,  $M$ is the mass of proton, $J_0$ and $J_1$  are cylindrical Bessel functions. The electric form factor $G_E$ and magnetic form factor $G_M$ of proton are given in the following parameterized form~\cite{Kelly:2002if,Alberico:2008sz,Arrington:2007ux} :
\begin{equation}
	G_E,\frac{G_M}{\mu_p}=\frac{1+\sum_{i=1}^na_i^{E,M}\tau^i}{1+\sum_{i=1}^{n+2}b_i^{E,M}\tau^i}.
\end{equation}
With these parameterization,  the plane charge density of a proton polarized along $+x$ direction is shown in Fig\ref{fig-rho2D-polarzied-p}. we can see that the center of charge moved along $-y$ direction.

\begin{figure}
		\includegraphics[width=0.8\linewidth]{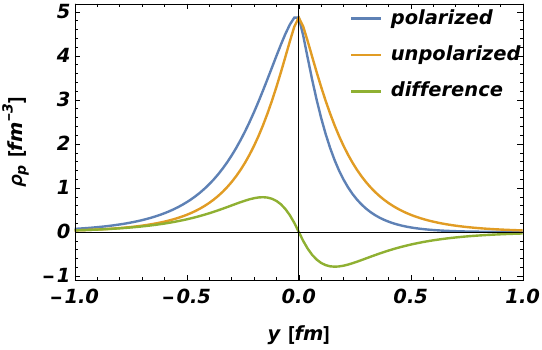}
	\caption{The 3-dimensional charge profiles in an unpolarized  proton (orange curve),  and a proton polarized along $+x$ (blue curve) as function of $y$ at $x=0$ and $z=0$.  }
	\label{fig-rho3D-compare}
\end{figure}

With this plane charge distribution in a polarized proton, we can derive its 3-dimensional charge profile using the same method in \cite{Zhang:2021jrc} . Shown in Fig\ref{fig-rho3D-compare},  we compare the 3-D charge profile of a polarized proton  and an unpolarized proton.  Their difference defined as $\rho_{pol} - \rho_{unpol}$, also shown in this figure. We can see that there is a significant violation to spherical symmetry in 3-dimensional charge profile as our expectation.

Using Lienard-Wiechert potential, we can calculate the spatial distribution of the electromagnetic field generated by a single flying polarized proton.
\begin{eqnarray}
	e \bm{E}{(t,\bm{r})}&=&\frac{e^2}{4\pi} \sum_{n}Z_n(\bm{R})\frac{\bm{R}_n-R_n\bm{v}_n}{(R_n-\bm{R}_n{\cdot}\bm{v}_n)^3}(1-\bm{v}_n^2)      \nonumber \\
	e\bm{B}{(t,\bm{r})}&=&\frac{e^2}{4\pi} \sum_{n}Z_n(\bm{R})\frac{\bm{v}_n \times \bm{R}_n}{(R_n-\bm{R}_n \cdot \bm{v}_n)^3}(1-\bm{v}_n^2)  \nonumber \\
\end{eqnarray}
here $Z_n$ is the $n^{th}$ particle charge number,  $v_n$  is its velocty. $R_n=r-r_n$ is the relative position of the field point $r$ and the source point $r_n$ which is the $n^{th}$ particle at the position with a delay time of $t_n=t-\left|r-r_n\right|$ \cite{Deng:2012pc}.

\begin{figure}
	\begin{tabular}{cc}
		\includegraphics[width=0.45\linewidth]{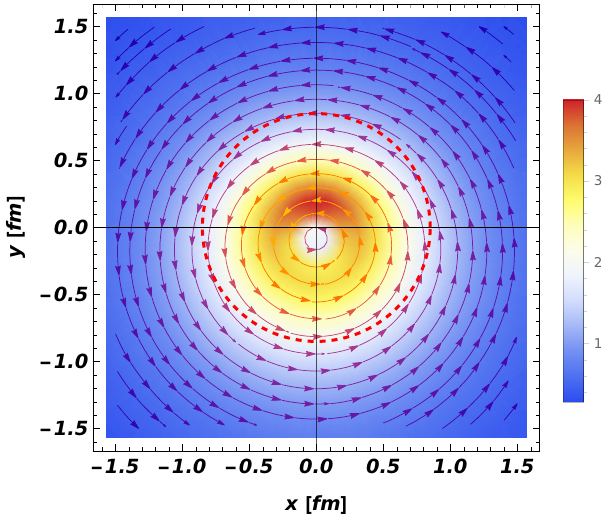} &    \includegraphics[width=0.45\linewidth]{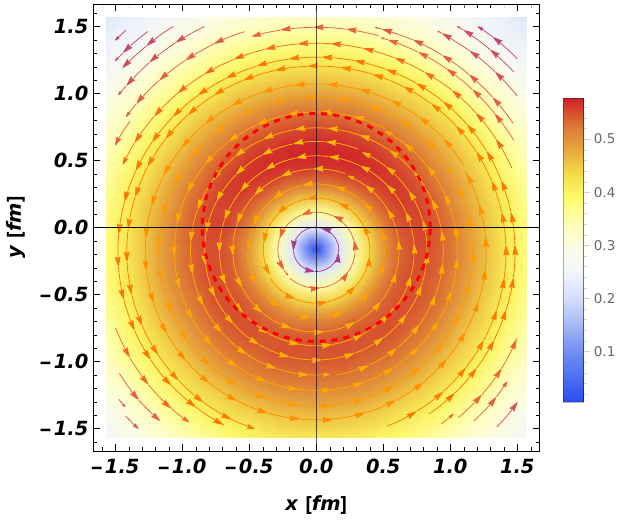}        \\
	\end{tabular}
	\caption{The distribution of magnetic field $\bm{B}$ on plane with $z=0$ (left panel), and $z=1/\gamma $fm (right panel) at $t=0$. A red circle with $r = 0.85 $fm is shown together, which is typical radii of proton.  }
			\label{fig-bvector-polP}
\end{figure}

We set the proton polarized along $+x$ axis still, and flying with a RHIC energy scale (100 GeV) along $+z$ direction. $t=0$ is set as the proton just flying through origin of coordinate $z=0$.  Shown in Fig\ref{fig-bvector-polP} is the magnetic field produced by this polarized proton at $t=0$.  Left panel of it showed the vector stream (vector) and strength (color) on plane with $z=0$, while right panel showed them on plane with $z=1/\gamma fm$ with a Lorentz factor $\gamma = E/m_p$.

\begin{figure}
	\begin{tabular}{cc}
		\includegraphics[width=0.45\linewidth]{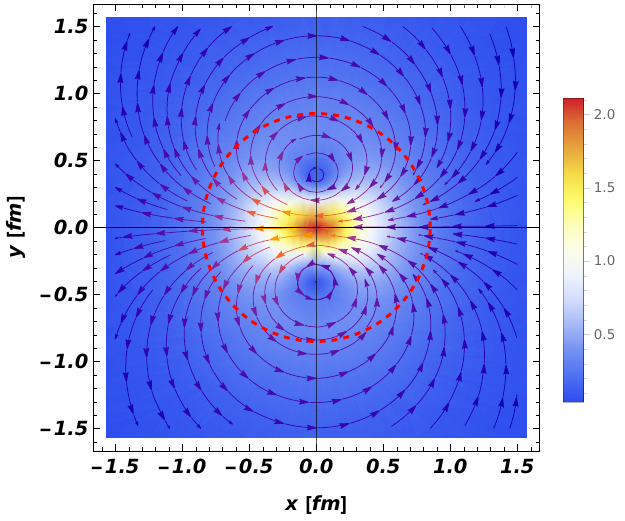} &    \includegraphics[width=0.45\linewidth]{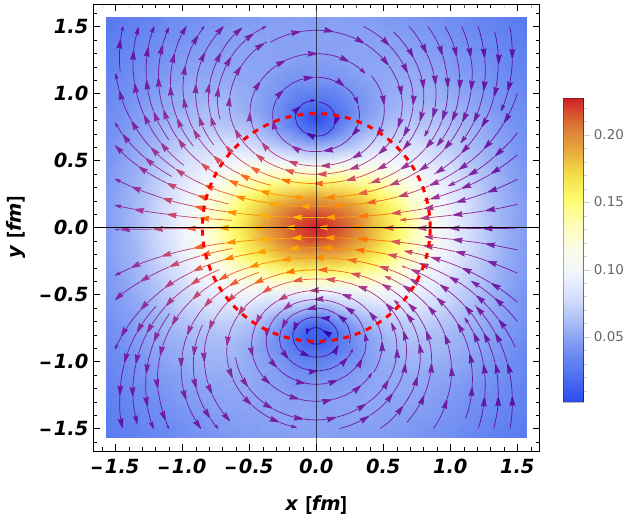}  
	\end{tabular}
	\caption{The distribution of  magnetic field difference $\Delta\bm{B}$ on plane with $z=0$ (left panel), and $z=1/\gamma $ fm (right panel) at $t=0$. A red circle with $r = 0.85 $ fm is shown together, which is typical radii of proton.  }
			\label{fig-bvectorDiff-polP}
\end{figure}

To compare the magnetic field produced by polarized  and unpolarized proton, we defined $\Delta \bm{B} = \bm{B}_{pol} - \bm{B}_{unpol}$ .  The results are shown in Fig\ref{fig-bvectorDiff-polP}. 
We can see that the  intensity of $\Delta\bm{B}$ is same order of the strength of $\bm{B}$ within the circle of   proton size, which will be the  overlap region in small collision system p+A.  

\begin{figure}
		\includegraphics[width=0.8\linewidth]{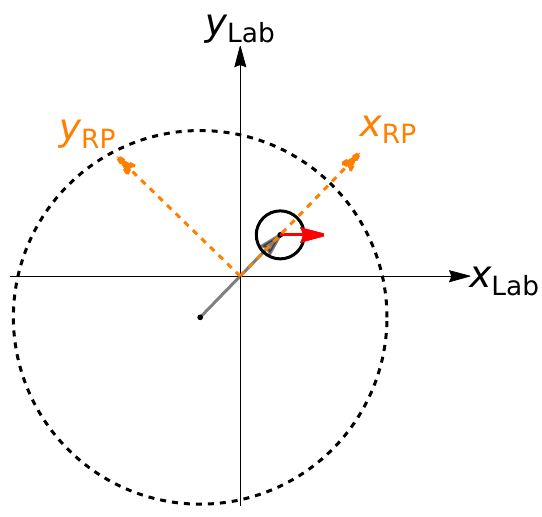} 
	\caption{An Illustration of  collision geometry of $p^\uparrow$ +Au with  $\Phi_{RP} \ne 0$.  }
	\label{fig-pA-geometry}
\end{figure}

\begin{figure}
	\center
	\begin{tabular}{cc}
		\includegraphics[width=4cm]{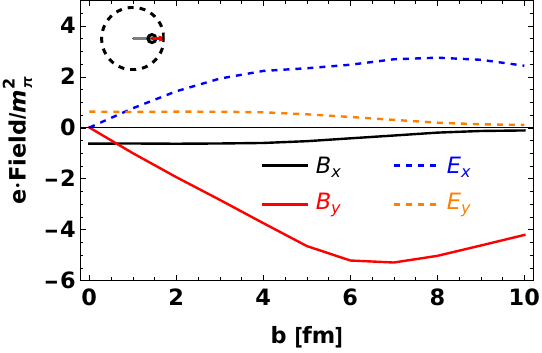} &    \includegraphics[width=4cm]{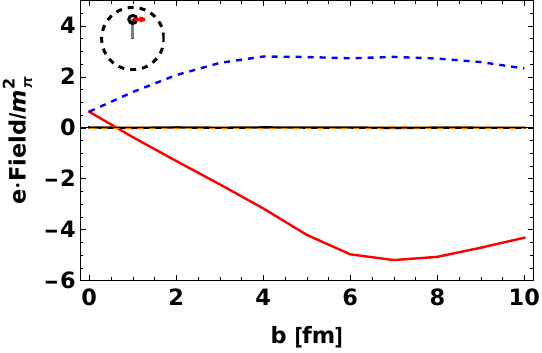}        \\
		(I) & (II) \\
		\includegraphics[width=4cm]{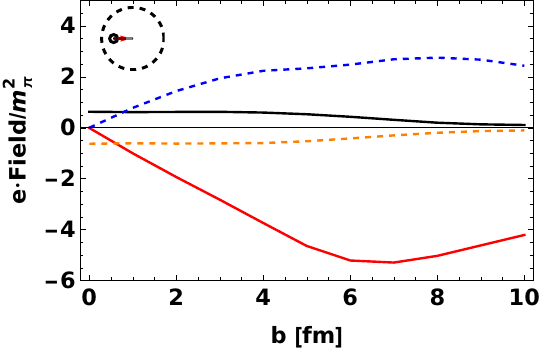} &    \includegraphics[width=4cm]{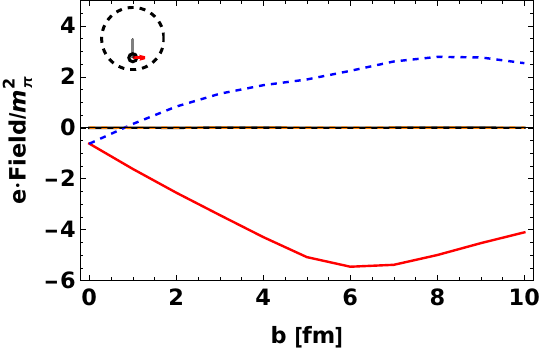}        \\
		(III) & (IV) \\
	\end{tabular}
	\caption{The strength of electromagnetic  fields  $B_x$, $B_y$, $E_x$, and $E_y$  as function of  impact parameter $b$ in four collision geometry schemes. }
	\label{fig-EMfields-4schemes}
\end{figure}

So, let's move from a flying single polarized proton to small collision system $p^\uparrow$ +Au  at RHIC energy $\sqrt{s}=200$GeV.  The collision geometry is set as the momentum of proton being $+z$ direction, while the Au nuclei being $-z$ direction. And the polarization is fixed as $+x$ direction. Impact parameter is defined as the vector point from center of Au to center of proton. Since the azimuthal angle of reaction plane $\Phi_{RP}$ is random in each event, we showed an illustration of one event with  $\Phi_{RP} \ne 0$ in Fig\ref{fig-pA-geometry} . 
In our following calculation, just four collision geometry schemes are investigated, with  $\Phi_{RP}=0$, $\pi/2$, $\pi$, and $3\pi/2$ respectively, illustrated in Fig\ref{fig-EMfields-4schemes}.

Then HIJING model \cite{Gyulassy:1994ew,Deng:2010mv} is employed to simulate the p+A collision process event by event in each collision scheme.  In our calculation of electromagnetic fields, the 3-D unsymmetrical charge profile  is used for the transversely  polarized projectile proton, and 3-D symmetrical charge profile is used for nucleons in target Au. 

Our results of electromagnetic fields on the center of overlap region averaged over events are shown in  Fig\ref{fig-EMfields-4schemes} for the four schemes. Components of $\bm{B}$ and $\bm{E}$ are projected into frame of event plane. So we can see that the value of $B_y$ is always negative as our expectation.

\begin{figure}
	\center
	\begin{tabular}{cc}
		\includegraphics[width=4cm]{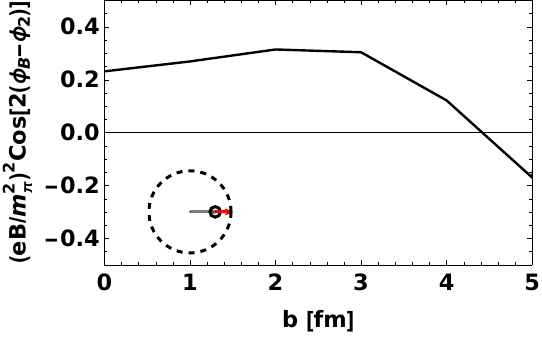} &    \includegraphics[width=4cm]{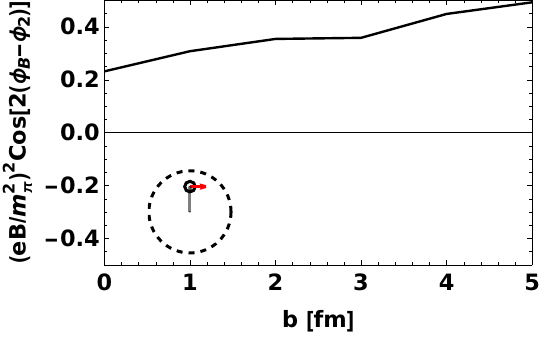}        \\
		(I) & (II) \\
		\includegraphics[width=4cm]{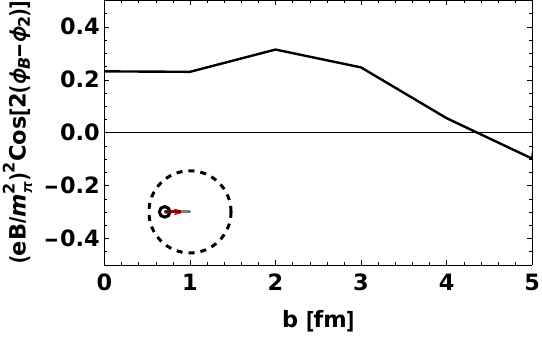} &    \includegraphics[width=4cm]{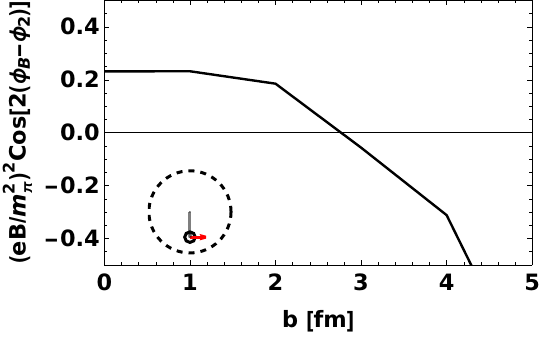}        \\
		(III) & (IV) \\
	\end{tabular}
	\caption{$B^2\cdot\cos [2(\Phi_B-\Phi_{2})]$ as function of  impact parameter $b$ in four collision geometry schemes. }
	\label{fig-Delta-gamma-4schemes}
\end{figure}

In order to get the CME signal  we calculated  correlator as Eq\ref{Eq-Delta-gamma-Phi2} , here $\Phi_2$ is the  azimuthal angle of event plane which is determined by all participants nucleons in each event\cite{Deng:2011at}.  Although the difference between $\Phi_{RP}$ and $\Phi_2$  is small,  the correlator as Eq\ref{Eq-Delta-gamma-Phi2} is more convenient to compare with possible future experiment data . 
\begin{equation}
	\Delta\gamma_\mathrm{CME} \propto B^2\cdot\cos [2(\Phi_B-\Phi_{2})],
	\label{Eq-Delta-gamma-Phi2}
\end{equation} 
The results of the correlators as function of impact parameter $b$ in the four schemes are shown in Fig\ref{fig-Delta-gamma-4schemes}. In each scheme, the correlator is not zero at small $b$ because of the fluctuation of $\bm{B}$ is not vanished in each event. Especially, we noticed that there is a significant difference between scheme II and scheme IV.  These two schemes are corresponding to the azimuthal angle of impact parameter $\bm{b}$ lying  on the $+y$ and $-y$ sectors respectively in lab frame. People can distinguish these two schemes in experiment with Zero Degree Calorimeters on RHIC.
\begin{figure}
	\includegraphics[width=0.8\linewidth]{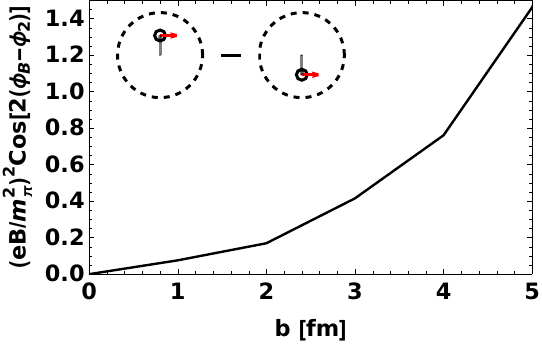} 
	\caption{Difference of $B^2\cdot\cos [2(\Phi_B-\Phi_{2})]$  between scheme II and scheme IV, as function of impact parameter $b$.  }
	\label{fig-Deltacorrelator-II-IV}
\end{figure}

It is more obvious to see the difference of CME correlators  between scheme II and IV $\Delta\gamma_\mathrm{CME~II} - \Delta\gamma_\mathrm{CME~IV}$, shown in Fig\ref{fig-Deltacorrelator-II-IV}. The  difference in thees two schemes increase with impact parameter, and reach a same order of  Au+Au collision.  So, if QGP is produced in small collision system, and CME exist here, its contribution to correlator will be much different between scheme II and scheme IV. On the other hand, background flow effects can be canceled completely even if it's not negligible when we check the difference between experiment data of scheme II and scheme IV.  
\begin{align}
	\nonumber
	\Delta\gamma_\mathrm{II} -  \Delta\gamma_\mathrm{IV}   \propto &
	 (\Delta\gamma_\mathrm{CME~II}  +  \Delta\gamma_\mathrm{BG~II} )  \\
	 \nonumber
	& - (\Delta\gamma_\mathrm{CME~IV}  +  \Delta\gamma_\mathrm{BG~IV} )  \\
	 \propto & \Delta\gamma_\mathrm{CME~II} -  \Delta\gamma_\mathrm{CME~IV}
	\label{Eq-Delta-gamma-II-IV} 
\end{align} 
So this $p^{\uparrow}+A$ small collision system can provide us an ideal  method to check the contribution of CME.

The authors acknowledge the support from the National Natural Science Foundation of China (No.  12075094). The computation is completed in the HPC Platform of Huazhong University of Science and Technology. We thank Prof. Qing-Hua Xu for his helpful discussion.
\bibliographystyle{apsrev4-2}
\bibliography{reference}

\end{document}